\documentclass[aps,pra,twocolumn,notitlepage,nofootinbib,superscriptaddress]{revtex4-1}

\usepackage{rahulbhadani_arxiv}
\usepackage{amsmath}
\usepackage{amssymb}
\usepackage{amsthm}
\newtheorem{definition}{Definition}
\usepackage{graphicx}
\usepackage[colorlinks=true,allcolors=blue]{hyperref}
\usepackage{braket}
\usepackage{dsfont}
\usepackage{multirow}
\usepackage{subcaption}
\usepackage{listings}
\begin{document}

\title{A Survey on Differential Privacy for SpatioTemporal Data in Transportation Research}

\author{Rahul Bhadani}
\email{rahul.bhadani@uah.edu}
\affiliation{The University of Alabama in Huntsville}


\begin{abstract}
With low-cost computing devices, improved sensor technology, and the proliferation of data-driven algorithms, we have more data than we know what to do with. In transportation, we are seeing a surge in spatiotemporal data collection. At the same time, concerns over user privacy have led to research on differential privacy in applied settings. In this paper, we look at some recent developments in differential privacy in the context of spatiotemporal data. Spatiotemporal data contain not only features about users but also the geographical locations of their frequent visits. Hence, the public release of such data carries extreme risks. To address the need for such data in research and inference without exposing private information, significant work has been proposed. This survey paper aims to summarize these efforts and provide a review of differential privacy mechanisms and related software. We also discuss related work in transportation where such mechanisms have been applied. Furthermore, we address the challenges in the deployment and mass adoption of differential privacy in transportation spatiotemporal data for downstream analyses.
\end{abstract}

\maketitle

\section{Introduction}
\label{sec:introduction}
In the age of information, we have multiple ways to obtain data representing human lives -- from the scale of the universe to the level of genomes. The proliferation of computing technology, sensor systems, and innovation for mass adoption provides timeseries data in IoT, healthcare, finance, log data for an event, transportation, genomics, and so on. Of particular interest to us is the spatiotemporal data. Spatiotemporal data are collected across both time and space, featuring at least one spatial and one temporal property~\cite{john2024spatiotemporal}. Spatiotemporal data may be trajectory data collected from vehicles such as speed, acceleration, brake, and throttle, along with their GPS information~\cite{bunting2022data}, weather data such as temperature pattern, precipitation, humidity, air quality, etc.~\cite{radhika2022novel}, species distribution mapping in ecology study~\cite{hengl2008advancing}, COVID-19 dataset~\cite{spassiani2021spatiotemporal}. Given that spatiotemporal data are timeseries data with spatial components, sequence-based methods have been used for analyses of spatiotemporal data such as Autoregressive Integrated Moving Average (ARIMA)) model~\cite{bouznad2020trend,yan2010traj,rodrigues2020spatiotemporal,li2022overview}, smoothing methods~\cite{rezaeiassessment,yang2018optimized}, and machine learning methods such as Long Short-Term Memory (LSTM)~\cite{alhudhaif2024spatio,luo2019spatiotemporal,li2021spatial}, and Transformers~\cite{abideen2020deep,ji2023traffic,yang2024stfeformer}. It is logistically challenging to gather or obtain spatiotemporal data specific to particular scenarios in the transportation domain due to the man-power required, lack of appropriate infrastructure, risks involved, and safety-critical nature of experiments~\cite{bhadani2018dissipation,stern2018dissipation,lee2024traffic,xu2022opv2v,schwall2020waymo}. Hence, many researchers rely on data gathered by other research groups, agencies, or companies for further analyses and secondary research. Releasing such spatiotemporal data may expose the private information of users and other stakeholders involved in the research, and may have repercussions beyond the experiment in the question. Developing some kind of privacy-preserving mechanism for such spatiotemporal is essential before the dissemination of such datasets can be done.

Differential privacy (DP)~\cite{dwork2006differential} is a class of statistical methods to prevent an attacker from identifying a particular person from the given dataset, possibly by associating with already available information~\cite{cowan2024handson}. Several differential privacy schemes have been proposed over the years suited for query-based data analysis from databases such as counting, sum, frequency~\cite{li2020estimating,wang2018privacy,yang2017survey} where privacy individuals are preserved when aggregated data are to be released. Researchers are also seeking to make their machine-learning models private by employing the DP mechanism in their machine-learning methods. In machine learning training, if no privacy-preserving mechanism is employed, it can compromise the confidentiality of the training data. A specific case of membership attack is discussed in~\cite{shokri2017membership,yeom2018privacy} where the attacker aims to determine whether a specific data record was used in the model’s training set. By consistently modifying these data points to increase the model's confidence, the actor can eventually pinpoint data samples that are nearly identical to the actual data samples in the training set. If a model is large, as in the case of a large language model, where there are millions to billions of parameters, the model will most likely memorize the training data~\cite{carlini2022quantifying}. In~\cite{carlini2021extracting,mattern2023membership}, a training data extraction attack is discussed where generative output can be engineered to reproduce the training data. Such cases were revealed from the output of some user-facing generative AI tools such as Midjourney~\cite{radauskas2024midjourney}. Violations of privacy with pre-trained machine learning models are possible and new mechanisms have been used for thwarting such violations. Some common work on the use of differential privacy in machine learning are Differentially Private Stochastic Gradient Descent (DP-SGD)~\cite{yu2022individual}, Private Feature Selection~\cite{stoddard2014differentially}, Private Classification~\cite{shen2023classification}, DP Neural Network~\cite{gylberth2017differentially} to name a few.

Despite some exciting developments in the use of DP methods in data analysis and machine learning, spatiotemporal data offers some specific challenges due to temporal relationships as well as spatial relationships among data points. We cannot directly use previously developed DP methods on spatiotemporal data as protecting every single data point, or removing some samples from the dataset may render the whole dataset useless. Correlation among features of multivariate spatiotemporal data may also lead to privacy violations if a well-thought method is not applied.

To the best of our knowledge, there is a lack of any review on differential privacy for spatiotemporal transportation data. We find surveys on spatiotemporal data mining, differential privacy in general, and differential privacy for timeseries. However, in the domain of transportation, no such work exists. We hope that this work will lead to renewed interest in applying differential privacy methods on transportation spatiotemporal data to aid researchers and users in performing private data analyses.

The rest of the paper is organized as follows: Section~\ref{sec:sp_analysis} introduces spatiotemporal data analysis to the reader. Section~\ref{sec:dp_primer} provides an overview of differential privacy. Section~\ref{sec:dp_sp} provides recent developments in the use of differential privacy on spatiotemporal data analyses. Finally, Section~\ref{sec:open_c} discusses open challenges and future directions in DP-based spatiotemporal data analyses.  

\section{SpatioTemporal Data}
\label{sec:sp_analysis}
Spatial data contains information about spaces and may contain data to represent objects like lines, circles, and polygons. Spatial data are common in environmental studies, geographical information systems (GIS),  and location-based services using global positioning systems (GPS). Spatial data may be map data, attribute data, or image data. Spatial data may be represented using coordinates and may be saved in databases as raster models or vector models. Of particular interest in transportation is HD maps. High-definition (HD) maps are highly detailed, three-dimensional representations of road environments used by automated vehicles (AVs) to navigate safely and efficiently. Unlike traditional maps, HD maps provide centimeter-level accuracy and include various layers of information critical for AV operation. They may contain information in the form of various layers such as a base layer; a geometric map containing 3D information of the road network; a semantic map layer containing information on lane markings, traffic signals,  and travel direction; and map prior layers such as crowd-sourced traffic data. It is used for autonomous navigation safely and reliably. Temporal data, also known as timeseries are data about representing a particular feature evolving over time. When temporal data is associated with spatial data, we obtain spatiotemporal data. For example, timeseries information about a vehicle's trajectory along a highway is spatiotemporal data. We may collect spatiotemporal data using mobile phones, social media, or sensors from moving vehicles.

\subsection{Analysis using SpatioTemporal Transportation Data}
Spatiotemporal data in transportation can be used to predict travel demand, travel time of transportation, traffic assignment problems, and route choices~\cite{yuan2021survey,xie2020urban}. In~\cite{zhang2017probe} addresses the problem of traffic congestion using accurate traffic forecasting using spatiotemporal data that was collected from probe vehicles. Authors use the Gray-level Co-occurrence Matrix (GLCM)~\cite{nixon2019feature} to extract spatiotemporal features. Using GLCM from current data and historical data, the authors calculated Normalized Squared Differences and selected best-matched traffic patterns~\cite{nixon2019feature}. In~\cite{liu2019contextualized}, authors address taxi origin-destination demand prediction using a spatiotemporal data-based neural network by adding context information -- spatial context, temporal context, and global correlation context. Authors consider taxi origin-destination demand in time interval $t$ as a 3D matrix $X_t \in \Rbb^{N\times H \times W}$, where $H$ and $W$ are height and width of city grid map respectively, $N$ is the total number of regions in the city. To predict the origin-destination demand, authors train on the New York City taxi dataset using a convolution-LSTM-based neural network. In~\cite{rahman2023data}, authors present a graph convolutional neural network (GCN)-based approach to learning traffic flow patterns by phrasing it as a traffic assignment problem. They present their transportation network as a graph which is a suitable data type for a graph-based neural network. Using GCN, they estimate the link flows for given travel demands. During the estimation process, the GCN model was able to learn the traffic flow propagation from origin nodes toward destination nodes in a transportation network. The authors validated their results by comparing the prediction against the analytical solution obtained from running static user equilibrium-based traffic assignments. Spatiotemporal data can also be used for collaborative filtering. Collaborative filtering employs the idea that people with similar spatiotemporal data may have similar preferences. This technique is used in creating a recommender system in transportation for people matching similar spatiotemporal features. For example, in~\cite{zhou2021smart}, authors create tourism recommendation system using spatiotemporal data.  Traditional methods often fail to consider real-time geographic conditions and tourists' dynamic needs, leading to suboptimal recommendations and tour routes. The authors propose a novel tourism recommendation model that combines cellular geospatial clustering with a multivariate weighted collaborative filtering algorithm. Authors use Multivariate Weighted Collaborative Filtering that enhances traditional collaborative filtering by incorporating multiple weighted factors that reflect tourists' preferences, such as budget, travel time, attraction popularity, tour purpose, and transportation modes. The algorithm calculates the similarity between tourists' feature attributes and historical data to provide personalized recommendations.

\section{A Primer on Differential Privacy}
\label{sec:dp_primer}
Differential privacy is a statistics-based formalization that measures the amount of privacy that may be leaked if a user decides to participate in an experiment resulting in some data collection about the user. Differential privacy promises the user that an individual's data does not significantly affect the output of data analysis, and the user is not at risk more than he would be if he decided not to participate in the experiment. By introducing randomized algorithms that ensure outputs are statistically similar regardless of individual data present, differential privacy mitigates risks from auxiliary information and linkage attacks.

\begin{figure}
    \centering
    \includegraphics[width=\linewidth]{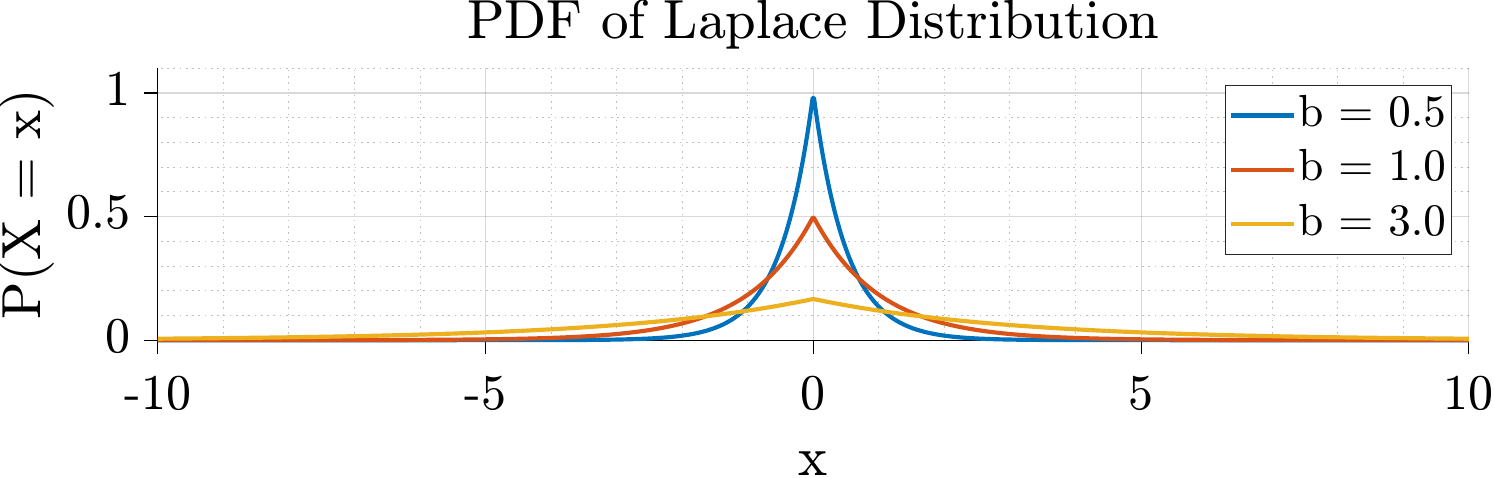}
    \caption{Probability Density Function (PDF) of the Laplace distribution with central parameter $\mu = 0$ and width parameters $b = 0.5$, $b = 1.0$, and $b = 3.0$.}
    \label{fig:LaplacePDF}
\end{figure}

One kind of randomized algorithm is the Laplace mechanism where a small amount of data drawn from the Laplace distribution is added to the dataset. Consider a generalized data transformation algorithm as $\Mcal$. Laplace mechanism can be applied as follows:
\begin{equation}
    \begin{aligned}
        \Mcal(x) = x + \textrm{Laplace(0, b)}
    \end{aligned}
\end{equation}
where $b$ is the width parameter of the Laplace distribution (See Figure~\ref{fig:LaplacePDF}). 

\begin{definition}{\textbf{$\epsilon$-Differential Privacy}}
    A data transformation $\Mcal$ is $\epsilon$-differentially private, if for a dataset $x$ and its perturbed dataset $x'$, we have the following condition:
    \begin{equation}
        P( \Mcal(x) = y) \leq P(\Mcal(x') = y) \cdot e^{\epsilon}
    \end{equation}
\end{definition}
which states that a transformation $\Mcal$ is $\epsilon$-DP if the probability (denoted by $P$) of any output $y$ from the original dataset $x$ is at most $e^\epsilon$ times the probability of the same output from a slightly perturbed dataset $x'$ that differs by only one data point. This ensures that the inclusion or exclusion of any single individual's data has a limited impact on the outcome, thus protecting individual privacy. In this formulation, $\epsilon$ is also referred to as privacy budget.

Considering the sensitivity $\Delta f$ of the function $f$ as 
$$
\Delta f = \max_{x, x'}||f(x) - f(x')||,
$$ the noise to be added can be sampled from

\begin{equation}
    \textrm{Laplace}\bigg(\cfrac{\Delta f}{\epsilon}\bigg)
\end{equation}

\begin{definition}{\textbf{$(\epsilon,\delta)$-Differential Privacy}}
    We consider $(\epsilon,\delta)$ model (called $(\epsilon,\delta)$-DP)  as
\begin{equation}
    P(\Mcal(x) = y) \leq e^{\epsilon} \cdot P(\Mcal(x') = y) + \delta
\end{equation}
\end{definition}

There are two models of differential privacy: the central model and the local model. In the central model, there is the existence of a trusted curator whose job is to apply some transformation $\Mcal$, after obtaining data from users or sources and then providing it for further analyses. In the local model, there is no trusted curator, hence, it is at the source, that we are required to apply specific transformation $\Mcal$. An illustration of both models is provided in Figure~\ref{fig:dp_model}.
\begin{figure}[h!]
    \centering
    \includegraphics[width=0.7\linewidth]{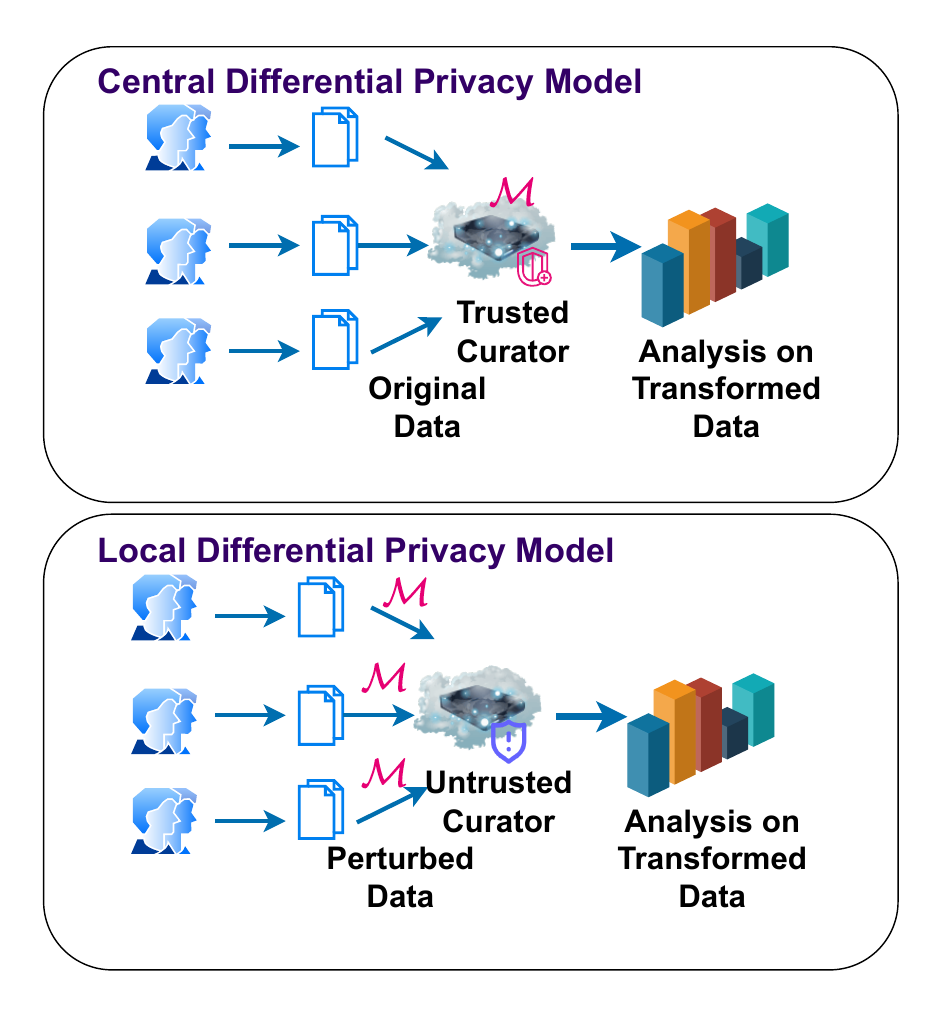}
    \caption{The central model and the local model for differentially private mechanism. A trusted curator collects data from its source, applies some transformation $\Mcal$, and disseminates it for downstream analyses. In the local model, the transformation is applied at the source as there may not be a trusted curator.}
    \label{fig:dp_model}
\end{figure}

\subsection{Algorithms for Differential Privacy}
Various algorithms exist for implementing differential privacy. Below, we discussed some of those algorithms:

\subsubsection{Randomized Response}
This algorithm is suitable for a case where a binary answer is possible. The respondent is instructed to use a randomization device, such as a coin flip or a random number generator, to determine whether they should provide their true answer or a randomized response. If the coin lands heads, the respondent answers truthfully. If the coin lands tails, the respondent answers \textit{Yes} with a probability $p$ (say, 0.5) and \textit{No} with a probability $1-p$. The surveyor collects the randomized responses without knowing whether any particular answer is truthful or randomized.

\subsubsection{Exponential Mechanism}
The exponential mechanism is used where adding noise might destroy the utility of the data. The Exponential Mechanism selects an output from a set of possible outcomes $r \in \Rcal$, $\Rcal$ being some arbitrary range, with a probability that exponentially favors outcomes with higher utility values, thus balancing the trade-off between utility and privacy.

Given a dataset $x$, a utility function $u(x, r)$ for each possible output $r\in \Rcal$ and a privacy parameter $\epsilon$, the exponential mechanism selects an output $r$ with probability proportional to $\exp{\epsilon u(x,r)/2 \Delta u}$ where $\Delta u$ is the sensitivity of the utility function defined as 

\begin{equation}
    \Delta u = \max_{r} \max_{x, x'} |u(x, r) - u(x', r)|
\end{equation}

As an example, Consider a scenario where a company wants to select the best location for a new store based on customer data while ensuring the decision process is differentially private. The utility function  $u(x, r)$ could represent the expected profit from placing the store at the location 
$r$ given the customer data $x$. The Exponential Mechanism would ensure that the chosen location does not reveal sensitive information about individual customers by making the probability of selecting each location proportional to its utility, scaled by the privacy parameter $\epsilon$.

\subsubsection{$\alpha$-net Mechanism}
$\alpha$-net mechanism is a technique used in the context of differential privacy to create privacy-preserving machine learning models. An $\alpha$-net is a carefully selected subset of the original dataset that aims to preserve the essential characteristics and patterns of the full dataset. The original dataset is divided into smaller, non-overlapping subsets. Each subset is analyzed to identify its key features and patterns. From each partition, a representative subset of data points is selected based on their utility and importance to the overall dataset. The selection process ensures that these points capture the essential characteristics of the partition while minimizing the risk of exposing individual data points. Controlled noise is added to the selected data points to ensure differential privacy. The representative subsets from each partition, collectively forming the $\alpha$-net, are used to train the machine learning model.

There are several other algorithms to provide guaranteed differential privacy that readers can find in the literature such as~\cite{hassan2019differential,yang2023local}.

\subsection{Software and Tools for Differential Privacy}
Research on differential privacy has led to many interesting software and tools that can be used to incorporate differential privacy into data analyses. We discuss some of the popular tools below.

\subsubsection{OpenDP}
OpenDP implements various statistical methods in rust with Python-binding~\cite{gaboardi2020programming}. OpenDP provides a flexible, extensible, and verifiable framework that can support a wide range of differential privacy algorithms and applications. The software introduces interactive measurements for adaptive queries and maintains privacy guarantees during post-processing. Additionally, it addresses implementation concerns such as preventing side-channel attacks and ensuring accurate arithmetic operations. The modular and extensible design aims to facilitate community contributions while maintaining rigorous privacy standards, providing a robust foundation for the advancement of differential privacy algorithms.

\subsubsection{Differential Privacy from Google}
Written primarily in C++, and Go, the Differential Privacy library from Google~\cite{GoogleDifferentialPrivacy2024} provides an implementation of various algorithms to introduce differential privacy to a data analysis task. The tool can also be used with SQL queries.

\subsubsection{TensorFlow Privacy}
Tensorflow-Privacy~\cite{papernot2019machine} is used to introduce privacy into machine learning models. It provides a framework for privacy mechanisms at the time of training a model that adds noise to the parameters of the model to prevent training data from being inferred by the model's output.

\subsubsection{SecretFlow}
SecretFlow~\cite{ma2023secretflow} provides a software framework for creating machine learning using secure multi-party computation (MPC) techniques. However, in addition to differential privacy, SecretFlow also offers other privacy-preserving techniques such as holomorphic encryption and Trusted Execution Environment.

\subsubsection{Opacus}
Opacus is a software package based on Pytorch for training deep-learning
models with differential privacy. It significantly simplifies the process of training differentially private models and achieves better performance compared to existing solutions. A developer can also track the privacy budget during the training process.

\section{Differential Privacy Methods on SpatioTemporal Data in Transportation}
\label{sec:dp_sp}
In this section, we discuss some of the recent work in spatiotemporal transportation data with differential privacy.

\subsection{Real-time Traffic Data Scenario}
In transportation, real-time traffic data publishing is desirable for congestion monitoring, crash detection, and connected vehicle control. In such a case, $w-event$ model~\cite{kellaris2014differentially} is a suitable choice.
Two spatiotemporal data $S$ and $S'$ are $w$-event adjacent if we consider a window of the size $w$ where the datasets are not identical, however, they are identical outside the window as illustrated in Figure.
\begin{figure}[h!]
    \centering
    \includegraphics[width=0.7\linewidth]{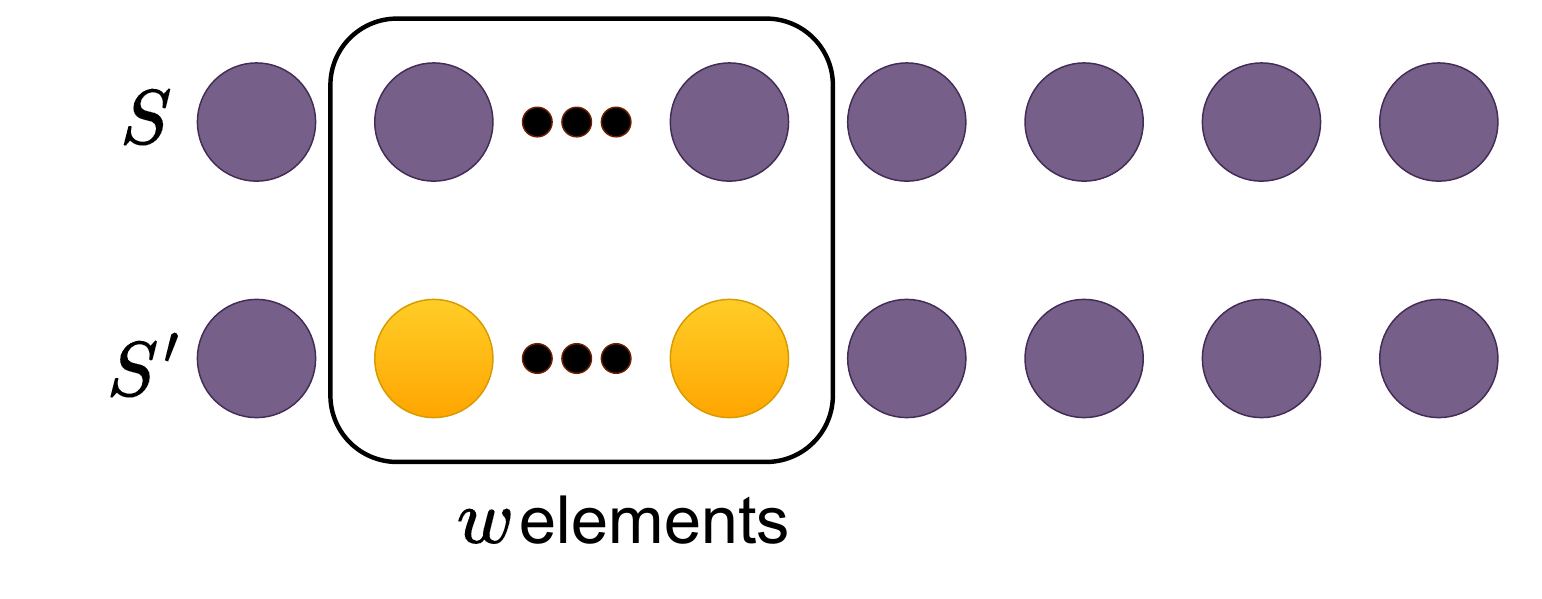}
    \caption{$w$-event model where two spatiotemporal data in a window of size $w$ are not identical.}
    \label{fig:dp_model}
\end{figure}
One key challenge in spatiotemporal data is how to allocate a privacy budget, i.e. choose $\epsilon$ when employing a sliding window. In addition, one should consider that observations made at nearby locations and timestamps may not be independent. There may be some correlation in temporal and spatial dimensions. Further, data distribution might change with time, leading to ineffective privacy budget allocation. In the context of transportation data, a dynamically allocated privacy budget might be used that takes into account distribution change over time and location.

In~\cite{wang2023dp}, authors propose to use a spatiotemporal graph attention network with a dynamic privacy budget (called P-STGAT). In P-STGAT, authors provide a distance-based adjacent matrix on road-network data. The adjacency matrix describes the magnitude of the correlation between nodes of the road network. The adjacency matrix is fed to a graph-attention network that contains a temporal attention layer and a spatial attention layer for extracting temporal and spatial features respectively. While calculating the new privacy budget, the last privacy budget is also taken into account, hence updating the privacy budget recursively as follows:
\begin{equation}
    \epsilon_{new} = \epsilon + \sum_{k = i - w + 1}^{i-1}\epsilon_k
\end{equation}
where $i$ is the final index in the current window of the size $w$, the $\epsilon$ is the fixed budget. Based on the allocated privacy budget, perturbed data using Laplace distribution is added to the original data.

\subsection{Trajectory Data Publication}
Trajectory data publication requires DP methods that take into account the fact that protection of only single-location data is sufficient as there are relationships between nearby location data. One solution is to apply differential privacy on processed data obtained from the true trajectory data. If we convert the original trajectory data into some other form such as latent representation (e.g. Principal Components), or low dimensional representation, then a single data point might contain information about more than one true data point.

In~\cite{qiu2023sgtp}, the authors first create K-means clusters based on the time of the trajectory positions, a process referred to as time generalization. After time generalization, all trajectories are generalized into $k$ temporal location sets and equal-length trajectories. In the next step, the density peak trajectory clustering algorithm (DPTC) is used to generalize spatial locations. DPTC assumes that cluster centers are data points surrounded by neighbors with lower local density and are relatively distant from any points with higher density. Once a clustered representation of the trajectory is obtained, Laplace distribution-based noise is added.

\subsection{Providing Spatiotemporal correlated Noise for Differential Privacy}
As the spatiotemporal data has a correlation in the spatial domain as well as the time domain, spatiotemporal noise is a suitable choice to prevent any privacy attack.

In~\cite{dou2020differential}, the authors use correlation functions to establish constraints on the published trajectory sequence, ensuring it maintains spatiotemporal correlation with both the original trajectory and the added noise. This is achieved by using the least squares method to fit the user's original trajectory and the overall direction of the noise sequence, constructing a noise candidate set that ensures the added noise sequence has a spatiotemporal correlation with the user's trajectory. Furthermore, the method imposes cross-correlation constraints to ensure the published trajectory sequence satisfies both temporal and spatial correlation constraints, thereby providing robust privacy protection.

The noise candidate set is constructed to ensure the spatial correlation between the user's original trajectory sequence and the added noise sequence. The least squares method is used to fit the direction of the original trajectory and the noise sequence as:
\begin{equation}
\hat{b} = \frac{\sum_{i=1}^n x_i y_i - n \bar{x} \bar{y}}{\sum_{i=1}^n x_i^2 - n \bar{x}^2} = \frac{\sum_{i=1}^n (x_i - \bar{x})(y_i - \bar{y})}{\sum_{i=1}^n (x_i - \bar{x})^2}
\label{eq:slope}
\end{equation}

where $\bar{x} = \frac{1}{n} \sum_{i=1}^n x_i$ and $\bar{y} = \frac{1}{n} \sum_{i=1}^n y_i$. To ensure the noise points are geographically reachable, the start and end points of the noise sequence need to  satisfy, 

\begin{equation}
|\text{dis}( \text{start}_d, \text{end}_d ) - \text{dis}( \text{start}_{\text{real}}, \text{end}_{\text{real}} )| \leq  \delta_{\text{dis}}
\end{equation}
and

\begin{equation}
|\text{time}( \text{start}_d, \text{end}_d ) - \text{time}( \text{start}_{\text{real}}, \text{end}_{\text{real}} )|\leq  \delta_{\text{time}}
\end{equation}

where $\delta_{\text{dis}}$ and $\delta_{\text{time}}$ are predefined thresholds for distance and time, respectively. As the calculation of the direction of the trajectory involves four data points (Equation~\eqref{eq:slope}), four groups $X_1, X_2, X_3, X_4$ of white Gaussian noise are sampled through Laplace distribution. Then $X = X_1^2 + X_2^2 - X_3^2 - X_4^2$ is then added to the original trajectory to achieve differential privacy.

\section{Open Challenges}
\label{sec:open_c}
While conducting the literature review for this paper, we found a lack of significant work on differential privacy specifically designed for spatiotemporal transportation data. There are interesting developments in differential privacy, as discussed in the previous sections. However, significant consideration is needed if we are to use spatiotemporal transportation data for making inferences or further data analyses.

First, there is generally a high degree of correlation in spatiotemporal data. Therefore, if a value is perturbed synthetically without considering spatial or temporal information, it may appear anomalous and be easily identified, undermining the principles of differential privacy. Second, spatiotemporal transportation data often capture high-dimensional data from various sensor modalities, which may be correlated across dimensions. Traditional methods might be insufficient in such scenarios. Finally, the current state of the art does not provide any evidence of successful autonomous vehicle operation using differentially private data. The existing literature lacks significant work on this topic.

A reasonable solution for the first two problems discussed above might be the release of synthetic data generated from true trajectory data using generative machine learning methods such as generative adversarial learning and diffusion-based methods. However, the feasibility of differential privacy for the autonomous control of vehicles is yet to be determined. There are real-time challenges in autonomous operation, as the real-time control of a vehicle requires control commands at a frequency of 20 Hz~\cite{hayat2023traffic}. An edge-based solution may be devised to meet the privacy budget in real-time.



 \bibliographystyle{splncs04}
\bibliography{references}

\end{document}